# Evaluation of simulation methods for tumor subclonal reconstruction

Jiaying Lai[1], Yunzhou Liu[1], Robert B. Scharpf[2,3], Rachel Karchin[1,2,3,4*]

[1]Institute for Computational Medicine, Johns Hopkins University, Baltimore, MD

[2]Sidney Kimmel Comprehensive Cancer Center, Johns Hopkins University School of Medicine, Baltimore, MD

[3]Department of Oncology, Johns Hopkins Medical Institutions, Baltimore, MD

[4]Department of Biomedical Engineering, Johns Hopkins University, Baltimore, MD

* Corresponding author

E-mail: karchin@jhu.edu

## Abstract

Most neoplastic tumors originate from a single cell, and their evolution can be genetically traced through lineages characterized by common alterations such as small somatic mutations (SSMs), copy number alterations (CNAs), structural variants (SVs), and aneuploidies. Due to the complexity of these alterations in most tumors and the errors introduced by sequencing protocols and calling algorithms, tumor subclonal reconstruction algorithms are necessary to recapitulate the DNA sequence composition and tumor evolution *in silico.* With a growing number of these algorithms available, there is a pressing need for consistent and comprehensive benchmarking, which relies on realistic tumor sequencing generated by simulation tools. Here, we examine the current simulation methods, identifying their strengths and weaknesses, and provide recommendations for their improvement. Our review also explores potential new directions for research in this area. This work aims to serve as a resource for understanding and enhancing tumor genomic simulations, contributing to the advancement of the field.

## Introduction

Tumors develop from sequential selection of cellular subpopulations from an initial progenitor cell (1, 2). The genetics of clonal evolution has been the subject of a large body of research, which has revolutionized our understanding of cancer. A variety of computational algorithms have been developed to reconstruct and model the hierarchical clonal architectures of tumors (3-44). Many algorithms are based on analysis of next-generation (NGS) tumor DNA sequence reads generated from bulk DNA or from single cells (scDNAseq). Clonal and subclonal markers derived from the reads include small somatic mutations (SSM), copy number alterations (CNA), structural variants (SV), and/or aneuploidies. With many algorithms now publicly available, the community needs datasets for benchmarking purposes. However, real tumor sequencing data does not provide ground truth about clonal evolution. Thus, simulating somatic alterations based on a known clonal structure has become a standard for evaluating subclonal reconstruction algorithms. For these simulations to be effective, they must accurately represent the complexities of actual tumor DNA (Fig 1).

We identified two major categories of simulators: 1) self-contained simulators that were designed for general use and were the subject of an independent publication; 2) custom simulators that were developed to benchmark a single group's reconstruction algorithm. We evaluated nine self-contained simulators: BAMSurgeon, MosaicSim, PSiTE, Pysubsim-tree, SCNVsim, Pysim-sv, CellCoal, SimSCSnTree, and CNAsim (45-53) (Table 1). Surprisingly, self-contained simulators were used for benchmarking by only two out of 42 published subclonal reconstruction algorithms (Table 2). For both the self-contained and custom simulators, we reported on their coverage of germline and somatic alteration types, models of clonal hierarchies, sequencing modalities, usability, support for more than one tumor sample or sequencing modality, and consideration of sequencing and caller errors.

## Few methods capture the complexity of tumor genome

Realistic tumor sequences include a mixture of reads from both tumor and normal DNA, covering various germline and somatic alterations on maternal or paternal chromosomes. Most methods initiated sequence generation with either a human reference genome FASTA file or a germline

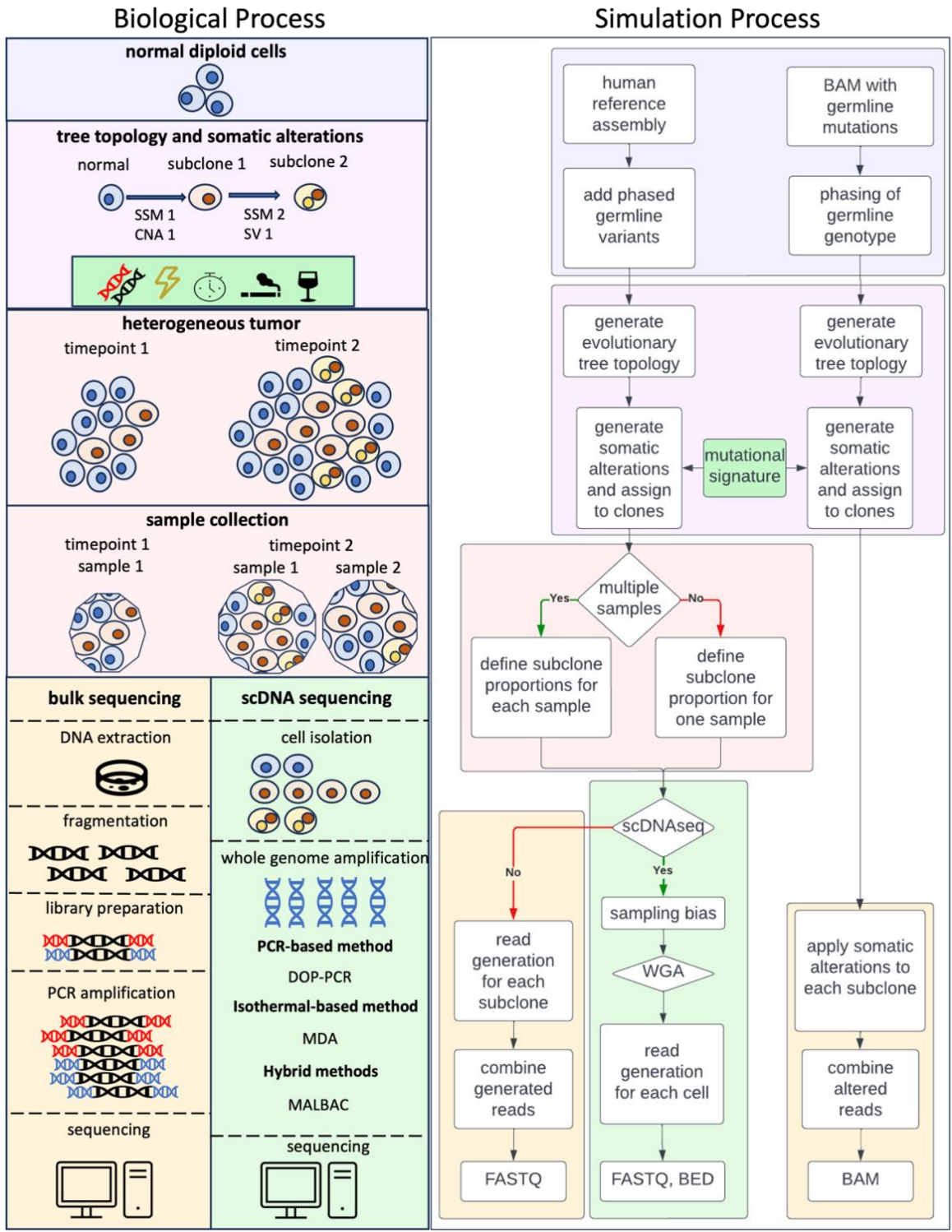

Figure 1. **Simulation methods for tumor subclonal reconstruction are designed to mimic biological tumor evolution**. Left panel: The process of tumor evolution begins when the DNA of a normal diploid cell acquires somatic alterations that transform it into a neoplastic cell. Through repeated mitosis, the originating cell generates a clone of genetically identical cells and continued

somatic alterations result in multiple subclones which proliferate or die at different rates. The evolution of (sub)clones is represented as a tree topology, with each (sub)clone assigned to a node. Tumors sampled at different time points may have varying subclone composition. Bulk DNA sequencing of tumor tissues requires DNA extraction and fragmentation, followed by library preparation in which adaptor sequences are ligated onto the fragments. For most technologies, the DNA is PCR amplified prior to sequencing. For scDNAseq, after cell isolation, most methods use whole-genome amplification to increase the amount of input material for sequencing, such as DOP-PCR, MDA, or MALBAC. Right panel: *In silico* simulation of tumor evolution begins with either a human reference assembly sequence or a germline BAM file, after which germline variants are added and/or phased to ensure a biologically realistic genome. Next, a tree topology is generated followed by simulated somatic alterations, including mutational signatures, which are assigned to (sub)clones on the tree. Each (sub)clone is assigned a proportion of the neoplastic cells in each sample. For simulation of bulk DNA sequencing, reads are generated for each subclone, then mixed to produce an output FASTQ file. For scDNAseq, reads are generated from each cell's simulated genome, and the simulated reads may correct for sampling bias or common single-cell sequencing errors, such as those resulting from WGA. The final output is either a FASTQ, BED, or BAM file for each simulated cell. SSM=small somatic mutation, CNA=copy number alteration, SV=structural variant, DOP-PCR=degenerate oligonucleotide-primed PCR, MDA=multi-displacement amplification, MALBAC= Multiple Annealing and Looping Based Amplification Cycles, WGA=whole-genome amplification, FASTQ=a text-based format to store raw reads, BED=Browser-Extensible Data format, BAM=Binary Alignment Map.

BAM file from a human subject (Fig 1). For the former, a list of germline and/or somatic alterations was bioinformatically spiked-in to the reference sequence. The reference sequence was then used to generate reads with a dedicated package, such as ART or Wessim (13, 15, 25, 33, 47, 49, 51-55), or was iteratively fragmented (46). If a germline BAM file was the starting point, somatic alterations were simulated by editing the reads (45). Both approaches produced FASTQ or BAM files of simulated tumor sequence. Alternate approaches simulated alterations to the reference sequence without generating reads (48) or produced high-level summaries of the sequence data such as read counts for SSMs, segmented read depth for CNAs, or a cell by mutation matrix for scDNAseq (Table 1 and 2). Of the evaluated methods, only three simulation methods can handle the full range of germline and somatic alterations in tumor genomes (45, 49, 53), highlighting a need for more sophisticated simulation techniques to capture the genetic diversity of tumors (Table 1).

## Assumptions about the process of tumor evolution vary

Simulation methods began with a model of an evolutionary process, which included a tree topology and a distribution of somatic alterations on the nodes or edges of the tree. For methods that assumed a coalescent process (46, 47, 50, 52), the topology was generated first, followed by the distribution of alterations. Alternatively, if a branching process was assumed (14, 16, 25, 37, 48, 49), the topology was generated in parallel with the alterations. Yet another approach was to generate a fixed or random topology and then generate the alterations, by parameterized distributions (Uniform, Poisson, Dirichlet etc.) (Table1 and 2). Four of the methods used coalescent models, but did not incorporate modifications for cancer evolution, such as selective sweeps (56-58). Another disadvantage of coalescent models was that they assumed a well-mixed population, which may be incorrect for solid tumors (56). Five methods iteratively applied a branching process to simulate either a clonal or cancer stem cell (CSC) model. Briefly, a clonal model was created by recursively generating subclones from a founder clone, and a CSC model by independently generating subclones from different founder clones (59, 60). However, these methods did not allow for consideration of differential clone fitness, such as parameterization of each clone with different birth and death rates. Finally, two self-contained methods required a user-provided tree with alterations assigned to the tree nodes (45, 53) (Table 1).

## Simulation for multiple tumor samples is limited

Most self-contained methods reviewed were geared towards simulating tumor sequences for a single sample, with some offering the option to randomly generate sequences for multiple samples. One unique method enabled sample generation at specific points within a three-dimensional space (47). While many custom simulation methods were capable of simulating bulk DNA data across multiple samples, none of these methods, whether self-contained or custom, accounted for the varying selective pressures that subclones might experience in different spatial locations or over time. This points to a need for improvements to encompass both spatial and temporal patterns in multiple tumor samples.

## Most methods lack targeted modalities

While whole-genome sequencing (WGS) has steadily become more affordable, cancer research increasingly utilizes targeted sequencing methods like whole-exome sequencing (WES) and gene panels for cost-effectiveness, especially at higher sequencing depths for better mutation detection. Simulating tumor sequences from WES or gene panels would greatly aid the benchmarking of subclonal reconstruction algorithms for these targeted approaches. However, among the reviewed methods that generate raw reads, only two supported targeted sequencing, relying on genomic region coordinates in BED format and, for one method, a probe definition file (46, 47). These methods allow customization of DNA fragment and read lengths, as well as read depth, with one method using an external package for sequence read simulation based on these parameters (47). Given the lack of targeted modalities in most methods, future developments should focus on enhancing the inclusion of targeted sequencing methods like WES and gene panels in simulation models.

## Error modeling for scDNAseq data is deficient

Single cell DNA sequencing identifies genetic information on a per-cell basis. In contrast to bulk sequencing, the mapping of reads to cells is observable, simplifying the process of subclonal reconstruction, but the results are error prone. Errors include sample distortion, false negative errors (coverage and allelic dropout, allelic imbalance), false positive errors (amplification errors, doublets) and missing values. Briefly, sample distortion occurs during the cell isolation step, in which cells are selected for sequencing based on size, viability, or likelihood of entering the cell cycle (61). Commonly used whole-genome amplification (WGA) methods prior to library construction, such as DOP-PCR, or MALBAC can introduce dropout, amplification errors, and substantially uneven coverage at different genomic locations (genome non-uniformity) (Fig 1). Three of the simulation methods generated single-cell reads, but did not incorporate any of these errors (46, 47). Two of the methods that generated reads incorporated genome non-uniformity (51, 52). Many methods incorporated multiple sources of error, but they generated cell-by-mutation matrix profiles rather than actual sequence data (Table 1 and 2). Sequencing errors confound subclonal reconstruction algorithms, and modelling errors in scDNAseq data simulations is crucial for accurate benchmarking. Advances like DLP Plus may reduce scDNAseq errors, potentially lessening the need for simulating these errors in the future (62). However, current error modeling in simulations must improve to match the technology's capabilities at present.

## Few methods offer combined bulk and scDNAseq simulation

Bulk DNA sequencing broadly detects somatic alterations in tumors but fails to link these to specific cells, while scDNAseq provides cell-specific details but is error-prone and may overlook tumor heterogeneity due to limited cell sampling. Combining these methods can harness their strengths. Seven of the methods we reviewed support both bulk and scDNAseq simulations for the same sample (Table1 and 2). We observed varied error handling capabilities in these methods, with some addressing multiple error types and others focusing on specific aspects like genomic non-uniformity. Enhancing current methods to include both, especially with refined scDNAseq error modeling, would be beneficial.

## Custom simulation methods predominate in benchmarking

We were surprised to discover that most subclonal reconstruction algorithms developed their own custom simulation methods for benchmarking purposes. Out of the 42 subclonal reconstruction algorithms (Table 2) only two used the software packages listed in Table 1: FastClone used DREAM challenge simulated tumor sequence from BAMSurgeon (17, 45, 63), and CellPhy used CellCoal (31, 50). The rest of the algorithms generated custom simulated data, which promoted less reproducible results and made unbiased comparisons between methods difficult.

## Less reproduciblility is seen in custom simulation methods

For scientific rigor, it is important that a benchmarking experiment can be reproduced by qualified researchers. Simulation reproducibility requires publishing either: 1) simulated sequence reads; 2) a high-level summary of the alterations; or 3) well-documented code that can generate sequence reads or a high-level summary, including necessary parameters. If the generation process includes randomization, a random seed must be included. If sequence reads are used, a well-documented, parameterized workflow for all steps of the analysis pipeline should be provided (genome alignment, germline and somatic alteration calling, as needed).

Thirteen of the subclonal reconstruction algorithms provided either the simulated sequence reads or the high-level summary of the data used in their benchmark experiments (13-15, 17, 18, 20, 22, 23, 26, 31, 38, 43, 44) (Table 2). The rest of the algorithms did not provide either type of benchmark data. Out of 42 algorithms, only eleven published their custom simulation code (Table 2). The quality of documentation ranged from no instruction about how to run a simulator (15, 29, 31, 32, 35, 38) to a simulator with a detailed usage guide (8, 13, 28, 34, 37). Seven of the methods also published the shell scripts or parameter files necessary to reproduce their benchmarking datasets (15, 28, 31, 34, 35, 37, 38). The findings highlight a need for more standardized, transparent, and replicable practices in the development and reporting of subclonal reconstruction algorithms to improve benchmarking in the field.

## Enhancements are needed for emerging technologies

As new sequencing technologies are developed and applied to analysis of tumor evolution, there are many opportunities for simulation methods to incorporate them. Long-read sequencers such as Sequel from Pacific Biosciences (PacBio) and minION from Oxford Nanopore Technologies (ONT) can now produce reads about 100-1500kb in length. This new generation of sequencers is enabling major improvements in phasing and SV calling, allowing identification of subclones with allele-specific alterations and SVs that have not been detectable with short-read technologies (64-67). Simulation methods can incorporate existing long-read sequence generators to produce long reads for benchmarking (68, 69). Another exciting direction is the emergence of *in situ* sequencing methods, in which each sequenced read includes a unique identifier (UID) and barcode representing its spatial location (70). Reads originating from a particular clone can be clustered and mapped to a spatial location on a tissue. This allows for observation of clone size, and when combined with single-cell transcriptomics, it is possible to identify the cell types and states present in each clone and its neighbors. *In situ* sequencing will play an important role in the next generation of subclonal reconstruction methods, and this will require new simulation methods that can generate realistic multi-modal *in situ* reads.

## Recommendations

1. **Standardize Simulation Methods:** To enable consistent and reproducible benchmarking of subclonal reconstruction algorithms, the field needs widely accepted protocols for how simulation tools should be used, which will simplify the process of comparing different algorithms. This can accelerate advancements in the field, as it becomes clearer what methods are most effective.

2. **Consider Complexity:** Aim to have simulation methods reflect a broad range of genomic alterations found in tumors, including phased germline and somatic alterations, where possible.

3. **Enhance Targeted Modality Support:** Develop simulations that support targeted

sequencing approaches like WES or gene panels to reflect current laboratory practices and clinical datasets.

4. **Refine Error Modeling:** Improve error modeling in simulations, particularly for single-cell DNA sequencing data, to reflect the artifacts present in modern sequencing technologies.

5. **Simulate Multi-sample Scenarios:** Extend simulation capabilities to cover multiple tumor samples and spatial locations, capturing intra-tumor heterogeneity and evolution.

6. **Document and Share:** Encourage comprehensive documentation and sharing of custom simulation codes and parameters to facilitate validation and replication of benchmarking results.

7. **Cross-Modality Simulation:** Promote the development of methods capable of simulating both bulk and single-cell DNA sequencing data to leverage their combined strengths.

8. **Community Collaboration:** Foster community-wide collaboration to refine and agree upon simulation standards and practices for the field. This ensures that the tools are not only used in a uniform manner but are also of a high standard.

## Conclusion

Evaluation of subclonal reconstruction algorithms requires comparison with ground truth about the evolutionary history of a neoplastic tumor. Currently, this ground truth cannot be measured experimentally, and realistic simulated data is required for benchmarking. Here we evaluated 51 simulation methods, considering their range of supported sequence alterations, sequencing modalities, assumptions about tumor evolution, reproducibility, and usability. We identified 9 methods that were designed for general use, but we discovered that most simulation methods were developed in conjunction with a particular reconstruction algorithm. Because the types of somatic alterations and assumptions about the tumor evolutionary process differed substantially among the simulation methods, it would be difficult to get stable results when benchmarking subclonal reconstruction algorithms with different simulators. The results of most custom simulation methods could not be reproduced, because code was either not published or not well documented. The diversity in methods and lack of reproducibility challenges effective benchmarking, highlighting the importance of developing more reliable and universally applicable simulation tools for tumor evolutionary studies.

# Acknowledgements

This study was funded by National Cancer Institute grant U01 CA271273 through the Cancer Systems Biology Consortium (CSBC) to (RK, RBS, JL).

**Table 1. Self-contained simulation methods.** The methods are distinguished by the types of sequence alterations they support **G**=germline, **S**=somatic, **GS**=germline and somatic, **SNV**=single nucleotide variant, **Indel**=small insertion/deletion, **CNA**=copy number alteration, **SV**=structural variant, **AP**=aneuploidy, **MS**=mutation signature, **LOH**=loss of heterozygosity. Details of supported sequencing modalities are **OT**=output type(=raw reads, =summary data such as reads counts for SNVs or segmented read depth for CNAs). Bulk sequencing features are **SE**=sequencing error, **BQ**=base quality, **GC**=GC bias, **WES**=whole-exome sequencing, **GP**=gene panel. scDNAseq errors are **FN**=false negative(such as allelic dropout and allelic imbalance), **FP**=false positives(such as amplification error and doublets), **MV**=missing value (result of non-uniform genome coverage), **SD**=sample distortion. The underlying evolutionary model is shown as a Tree type **UI**=user-input, **BP**=branching process, **CA**=coalescent, **RT**=random topology. **NMSS**=Native multi-sample support=number of samples is an input to the method. **DD**=detailed documentation.

| Methods | Genome alterations ||||||  Sequencing modalities |||||||||||| Tree | NMSS | DD | Ref |
| --- | --- | --- | --- | --- | --- | --- | --- | --- | --- | --- | --- | --- | --- | --- | --- | --- | --- | --- | --- | --- | --- |
|  |  |  |  |  |  |  | bulk DNA sequencing |||||| scDNAseq ||||| | | | |
|  | SNV | Indel | CNA | SV | AP | MS | OT | SE | BQ | GC | WES | GP | OT | FN | FP | MV | SD | | | | |
| BAMSurgeon | GS | GS | GS | GS | ✓ | ✓ |  | ✓ | ✓ | ✓ | - | - | - |  |  |  | - | UI | - |  | [45] |
| Pysubsim-tree | GS | GS | GS | GS | ✓ | - |  | ✓ | ✓ | ✓ | - | - | - |  |  |  | - | UI | - |  | [53] |
| Pysim-sv | GS | GS | GS | GS | ✓ | - |  | ✓ | ✓ | ✓ | - | - | - |  |  |  | - | BP | - |  | [49] |
| MosaicSim | S | S | S | S | ✓ | ✓ |  | ✓ | - | - | ✓ | ✓ |  |  |  |  | - | CA | ✓ | - | [46] |
| PSiTE | GS | GS | S | - | ✓ | - |  | ✓ | ✓ | ✓ | ✓ | ✓ |  |  |  |  | - | CA | ✓ |  | [47] |
| SimSCSnTree | S | - | S | - | ✓ | - |  | ✓ | ✓ | ✓ | - | - |  |  |  | ✓ |  | RT | - |  | [51] |
| CellCoal | GS | GS | GS LOH |  |  | ✓ | - | - | - | - | - | - |  | ✓ | ✓ | ✓ |  | CA | - |  | [50] |
| CNAsim | - | - | S | - | ✓ | - | - | - | - | - | - | - |  |  |  | ✓ |  | CA | - |  | [52] |
| SCNVsim | G | G | S | S | ✓ | - | - | - | - | - | - | - | - |  |  |  | - | BP | - |  | [48] |

**Table 2. Simulation methods used to benchmark 42 subclonal reconstruction algorithms.** Algorithms that used a self-contained method from Table 1 are indicated in bold type. Forty methods were self-benchmarked by a custom simulation method. **G**=germline, **S**=somatic, **GS**=germline and somatic, **SNV**=single nucleotide variant, **Indel**=small insertion/deletion, **CNA**=copy number alteration, **SV**=structural variant, **AP**=aneuploidy, **MS**=mutation signature. Details of supported sequencing modalities are **OT**=output type(🧬=raw reads, 📕=summary data such as reads counts for SNVs or segmented read depth for CNAs). Bulk sequencing features are **SE**=sequencing error, **BQ**=base quality, **GC**=GC bias, **WES**=whole-exome sequencing, **GP**=gene panel. scDNAseq errors are **FN**=false negative(such as allelic dropout and allelic imbalance), **FP**=false positives(such as amplification error and doublets), **MV**=missing value (result of non-uniform genome coverage), **SD**=sample distortion. **MSS**=multiple sample simulated , **Data**=simulated data used for benchmarking, and **Code**=simulation code. 📖=code includes detailed documentation, ⭐=parameters or shell scripts to reproduce simulation. Tree types are: **BP**=branching process, **CA**=coalescent, **RT**=random topology, or **EX**=exhaustive for fixed N, where N is the number of nodes on the tree. *=pre-computed trees were provided, and the number of trees is indicated.

| Reconstruction Algorithm | Custom Simulation Methods | | | | | | | | | | | | | | | | | Tree | MSS | Data | Code | Ref |
|---|---|---|---|---|---|---|---|---|---|---|---|---|---|---|---|---|---|---|---|---|---|---|
| | Genome alterations | | | | | | Sequencing modalities | | | | | | | | | | | | | | | |
| | | | | | | | bulk DNA sequencing | | | | | | scDNAseq | | | | | | | | |
| | SNV | Indel | CNA | SV | AP | MS | OT | SE | BQ | GC | WES | GP | OT | FP | FN | MV | SD | | | | | |
| **CellPhy** | S | - | - | - | - | - | - | - | - | - | - | - | 📕 | ✓ | ✓ | ✓ | - | CA | - | ✓ | ✓⭐ | [31] |
| **FastClone** | GS | GS | GS | GS | ✓ | ✓ | 🧬 | ✓ | ✓ | ✓ | - | - | - | - | - | - | - | 8* | - | ✓ | - | [17] |
| SVClone | G | - | - | S | - | - | 🧬 | ✓ | ✓ | ✓ | - | - | - | - | - | - | - | 1* | - | ✓ | ✓⭐ | [15] |
| SPhyR | S | - | - | - | - | - | - | - | - | - | - | - | 📕 | ✓ | ✓ | - | - | RT | - | ✓ | ✓⭐ | [38] |
| HATCHet | G | - | S | - | ✓ | - | 🧬 | ✓ | ✓ | ✓ | - | - | - | - | - | - | - | RT | ✓ | ✓ | ✓📖 | [13] |
| CNTMD | - | - | S | - | - | - | 📕 | - | - | - | - | - | - | - | - | - | - | RT | ✓ | ✓ | - | [44] |
| DEVOLUTION | - | - | S | - | - | - | 📕 | - | - | - | - | - | - | - | - | - | - | BP | ✓ | ✓ | - | [14] |
| SPRUCE | S | - | S | - | - | - | 📕 | - | - | - | - | - | - | - | - | - | - | RT | ✓ | ✓ | - | [20] |
| PhyloWGS | S | - | S | - | - | - | 📕 | - | - | - | - | - | - | - | - | - | - | RT | - | ✓ | - | [22] |
| SubMARine | S | - | S | - | - | - | 📕 | - | - | - | - | - | - | - | - | - | - | RT | ✓ | ✓ | - | [23] |
| TUSV-ext | S | - | S | S | - | - | 📕 | - | - | - | - | - | - | - | - | - | - | 1* | ✓ | ✓ | - | [26] |
| ddClone | S | - | S | - | - | - | 📕 | - | - | - | - | - | 📕 | ✓ | ✓ | - | ✓ | RT | - | ✓ | - | [43] |
| OncoNEM | S | - | - | - | - | - | - | - | - | - | - | - | 📕 | ✓ | ✓ | ✓ | - | RT | - | - | ✓📖⭐ | [28] |
| SCG | S | - | - | - | - | - | - | - | - | - | - | - | 📕 | ✓ | ✓ | ✓ | - | RT | - | - | ✓📖⭐ | [34] |
| SIEVE | GS | S | S | - | - | - | - | - | - | - | - | - | 📕 | ✓ | ✓ | ✓ | - | BP | - | - | ✓📖⭐ | [37] |
| SCIΦ | S | - | S | - | - | - | - | - | - | - | - | - | 📕 | ✓ | ✓ | ✓ | - | RT | - | - | ✓⭐ | [35] |

| Method | C1 | C2 | C3 | C4 | C5 | C6 | C7 | C8 | C9 | C10 | C11 | C12 | C13 | C14 | C15 | C16 | C17 | C18 | C19 | Ref |
|---|---|---|---|---|---|---|---|---|---|---|---|---|---|---|---|---|---|---|---|---|
| QuantumClone | S | - | - | - | ✓ | - | 📕 | - | - | - | - | - | - | - | - | - | RT | ✓ | - | ✓📖 | [8] |
| CONET | - | - | S | - | - | - | - | - | - | - | - | - | 📕 | - | - | - | - | RT | - | - | ✓ | [32] |
| SBMClone | S | - | - | - | - | - | - | - | - | - | - | - | 📕 | - | - | ✓ | - | 2* | - | - | ✓ | [29] |
| CITUP | S | - | - | - | - | - | 📕 | - | - | - | - | - | - | - | - | - | - | RT | ✓ | - | - | [3] |
| Rec-BTP | S | - | - | - | - | - | 📕 | - | - | - | - | - | - | - | - | - | - | EX | - | - | - | [4] |
| PhyloSub | S | - | - | - | - | - | 📕 | - | - | - | - | - | - | - | - | - | - | 1* | - | - | - | [5] |
| AncesTree | S | - | - | - | - | - | 📕 | - | - | - | - | - | - | - | - | - | - | RT | ✓ | - | - | [6] |
| TrAp | S | - | - | - | - | - | 📕 | - | - | - | - | - | - | - | - | - | - | RT | - | - | - | [7] |
| ClonosGP | S | - | - | - | - | - | 📕 | - | - | - | - | - | - | - | - | - | - | RT | ✓ | - | - | [9] |
| SCHISM | S | - | - | - | - | - | 📕 | - | - | - | - | - | - | - | - | - | - | EX | ✓ | - | - | [10] |
| PICTograph | S | - | - | - | - | - | 📕 | - | - | - | - | - | - | - | - | - | - | EX | ✓ | - | - | [11] |
| THetA | - | - | S | - | - | - | 📕 | - | - | - | - | - | - | - | - | - | - | 1* | - | - | - | [12] |
| LICHeE | S | - | S | - | - | - | 📕 | - | - | - | - | - | - | - | - | - | - | BP | ✓ | - | - | [16] |
| PyClone | S | - | S | - | - | - | 📕 | - | - | - | - | - | - | - | - | - | - | 1* | - | - | - | [18] |
| DeCiFer | S | - | S | - | ✓ | - | 📕 | - | - | - | - | - | - | - | - | - | - | RT | ✓ | - | - | [19] |
| Canopy | S | - | S | - | - | - | 📕 | - | - | - | - | - | - | - | - | - | - | 4* | ✓ | - | - | [21] |
| cloneHD | GS | - | S | - | - | - | 📕 | - | - | - | - | - | - | - | - | - | - | 1* | ✓ | - | - | [24] |
| MELTOS | S | - | S | S | - | - | 🧬 | ✓ | ✓ | ✓ | - | - | - | - | - | - | - | BP | ✓ | - | - | [25] |
| BnpC | S | - | - | - | - | - | - | - | - | - | - | - | 📕 | ✓ | ✓ | ✓ | - | RT | - | - | - | [27] |
| SCITE | S | - | - | - | - | - | - | - | - | - | - | - | 📕 | ✓ | ✓ | ✓ | - | RT | - | - | - | [30] |
| Ginkgo | - | - | S | - | - | - | - | - | - | - | - | - | 🧬 | - | - | - | - | 1* | - | - | - | [33] |
| SiFit | S | S | S | - | - | - | - | - | - | - | - | - | 📕 | ✓ | ✓ | ✓ | - | RT | - | - | - | [36] |
| SiCloneFit | S | - | S | - | - | - | - | - | - | - | - | - | 📕 | ✓ | ✓ | ✓ | - | RT | - | - | - | [39] |
| Conifer | S | - | S | - | - | - | 📕 | - | - | - | - | - | 📕 | ✓ | ✓ | ✓ | ✓ | RT | - | - | - | [40] |
| PhISCS | S | - | - | - | - | - | 📕 | - | - | - | - | - | 📕 | ✓ | ✓ | ✓ | - | RT | - | - | - | [41] |
| B-SCITE | S | - | S | - | - | - | 📕 | - | - | - | - | - | 📕 | ✓ | ✓ | ✓ | ✓ | RT | - | - | - | [42] |